\documentclass[aps,prl,10pt,twocolumn,nofootinbib,superscriptaddress,preprintnumbers]{revtex4}
\usepackage{slashed}
\usepackage{graphicx}
\usepackage{times}
\usepackage{epsfig}
\usepackage{amsmath}
\usepackage{amsfonts}
\usepackage{amssymb}
\usepackage{url}
\usepackage{hyperref}
\usepackage{subfigure}
\newcommand{\be}{\begin{equation}}
\newcommand{\ee}{\end{equation}}
\newcommand{\bea}{\begin{eqnarray}}
\newcommand{\eea}{\end{eqnarray}}

\begin{document}
\pagestyle{plain}
\title{A solution to the $\mu$ problem in the supersymmetric unparticle physics}
\author{Tatsuru Kikuchi}
\email{tatsuru@post.kek.jp}
\affiliation{Theory Division, KEK,
1-1 Oho, Tsukuba, 305-0801, Japan.}
\date{\today}
\begin{abstract}
Recently, conceptually new physics beyond the Standard Model 
 has been proposed by Georgi, where a new physics sector 
 becomes conformal and provides ``unparticle'' 
 which couples to the Standard Model sector through higher 
 dimensional operators in low energy effective theory. 
Among several possibilities, we focus on operators 
 involving the (scalar) unparticle, Higgs and the gauge bosons. 
Once the Higgs develops the vacuum expectation value (VEV), 
 the conformal symmetry is broken and as a result, 
 the mixing between the unparticle and the Higgs boson emerges.  
In this paper, we consider the unparticle as a hidden sector of supersymmetry (SUSY) breaking,
and give some phenomenological consequences of this scenario.
The result shows that there is a possibility for the unparticle as a hidden sector in
SUSY breaking sector, and can provide a solution to the $\mu$ problem
in SUSY models.
\end{abstract}
\maketitle
\section{Introduction}
In spite of the success of the Standard Model (SM) 
 in describing all the existing experimental data, 
 the Higgs boson, which is responsible for the electroweak 
 symmetry breaking, has not yet been directly observed, 
 and is one of the main targets 
 at the CERN Large Hadron Collider (LHC). 
At the LHC, the main production process of Higgs boson is 
 through gluon fusion, and if Higgs boson is light, say 
 $m_h \lesssim 150$ GeV, the primary discovery mode 
 is through its decay into two photons. 
In the SM, these processes occur only at the loop level 
 and Higgs boson couples with gluons and photons very weakly. 

A certain class of new physics models includes 
 a scalar field which is singlet under the SM gauge group. 
In general, such a scalar field can mix with the Higgs boson 
 and also can directly couple with gluons and photons 
 through higher dimensional operators 
 with a cutoff in effective low energy theory.   
Even if the cutoff scale is very high, say, 100-1000 TeV, 
 the couplings with gluons and photons can be comparable 
 to or even larger than those of the Higgs boson 
 induced only at the loop level in the SM. 
This fact implies that if such a new physics exists, 
 it potentially has an impact on Higgs boson phenomenology 
 at the LHC. 
In other words, such a new physics may be observed together 
 with the discovery of Higgs boson. 

As one of such models, in this letter, 
 we investigate a new physics recently proposed 
 by Georgi \cite{Georgi:2007ek}, which is described in terms 
 of "unparticle" provided by a hidden conformal sector 
 in low energy effective theory. 
A concrete example of unparticle staff was proposed by Banks-Zaks \cite{Banks:1981nn} 
many years ago, where providing a suitable number of massless fermions, 
 theory reaches a non-trivial infrared fixed points 
 and a conformal theory can be realized at a low energy. 
Various phenomenological considerations on the unparticle physics 
 have been developed in the literature \cite{U-pheno}. 
It has been found that inclusion of the mass term for the unparticle plays an important role especially 
in studying about the Higgs-unparticle systems \cite{U-Higgs}, indeed we have studied the unparticle
physics focusing on the Higgs phenomenology including the effects of the conformal symmetry breaking \cite{Kikuchi}, 
and there are some other studies on the Higgs phenomenology in the literature of the unparticle physics \cite{U-Higgs2}.
Inclusion of such effects of the conformal symmetry breaking or the infrared (IR) cutoff is also considered in the literature of 
hadron collider physics \cite{Rizzo}, and in the model of colored unparticles \cite{U-color}.
There has also been studied on the astrophysical and cosmological applications of the unparticle physics \cite{U-astro},
especially, we have proposed a possibility for the unparticle dark matter scenario \cite{UDM}.
And there are some studies on the more formal aspects of the unparticle physics \cite{U-formal}
and its effects to the Hawking radiation \cite{Dai:2008qn}.

Now we begin with a review of the basic structure of the unparticle physics. 
First, we introduce a coupling between the new physics operator 
 ($\cal{O}_{\rm UV}$) with dimension $d_{\rm UV}$ 
 and the Standard Model one (${\cal O}_{\rm SM}$) with dimension $n$, 
\bea
 {\cal L} = \frac{c_n}{M^{d_{\rm UV}+n-4}} 
     \cal{O}_{\rm UV} {\cal O}_{\rm SM} ,  
\eea
where $c_n$ is a dimension-less constant, and $M$ is the energy scale 
 characterizing the new physics. 
This new physics sector is assumed to become conformal 
 at a energy $\Lambda_{\cal U}$, and 
 the operator $\cal{O}_{\rm UV}$ flows to the unparticle operator 
 ${\cal U}$ with dimension $d_{\cal U}$. 
In low energy effective theory, we have the operator of the form, 
\bea
{\cal L}=c_n 
 \frac{\Lambda_{\cal U}^{d_{\rm UV} - d_{\cal U}}}{M^{d_{\rm UV}+n-4}}   
 {\cal U} {\cal O}_{\rm SM} 
\equiv 
  \frac{1}{\Lambda^{d_{\cal U}+ n -4}}  {\cal U} {\cal O}_{\rm SM},  
\eea 
where the dimension of the unparticle ${\cal U}$ have been 
 matched by $\Lambda_{\cal U}$ which is induced 
 the dimensional transmutation, 
 and $\Lambda$ is the (effective) cutoff scale of 
 low energy effective theory. 
In this paper, we consider only the scalar unparticle. 

It was found in Ref.~\cite{Georgi:2007ek}
 that, by exploiting scale invariance of the unparticle, 
 the phase space for an unparticle operator 
 with the scale dimension $d_{\cal U}$ and momentum $p$ 
 is the same as the phase space for 
 $d_{\cal U}$ invisible massless particles, 
\begin{eqnarray}
d \Phi_{\cal U}(p) = 
 A_{d_{\cal U}} \theta(p^0) \theta(p^2)(p^2)^{d_{\cal U}-2} 
 \frac{d^4p}{(2\pi)^4} \,,
\label{Phi}
\end{eqnarray}
where
\begin{eqnarray}
A_{d_{\cal U}} = \frac{16 \pi^{\frac{5}{2}}}{(2\pi)^{2 d_{\cal U}}}
\frac{\Gamma(d_{\cal U}+\frac{1}{2})}{\Gamma(d_{\cal U}-1) 
\Gamma(2 d_{\cal U})}.
\label{A}
\end{eqnarray}
%
Also, based on the argument on the scale invariance, 
 the (scalar) propagator for the unparticle was suggested to be \cite{U-propagator}
\begin{eqnarray}
 \frac{A_{d_{\cal U}}}{2\sin(\pi d_{\cal U})}
 \frac{i}{(p^2)^{2-d_{\cal U}}} 
 e^{-i (d_{\cal U}-2) \pi}  \,.
\label{propagator}
\end{eqnarray}
%
Because of its unusual mass dimension, 
 unparticle wave function behaves as 
 $\sim  (p^2)^{(d_{\cal U}-1)/2}$ (in the case of scalar unparticle).

\section{Unparticle and the Higgs sector}
First, we begin with a brief review of our previous work on
the Higgs phenomenology in the unparticle physics \cite{Kikuchi}.
Among several possibilities, we will focus on the operators 
 which include the unparticle and the Higgs sector, 
\be 
 {\cal L} = \frac{1}{\Lambda^{d_{\cal U}+ n - 4}} 
 {\cal U} {\cal O}_{\rm SM}(H^\dagger H)  
+\frac{1}{\Lambda^{2 d_{\cal U} +n -4}} 
{\cal U}^2 {\cal O}_{\rm SM}(H^\dagger H)  \;,
\ee
where $H$ is the Standard Model Higgs doublet and  
 ${\cal O}_{\rm SM}(H^\dagger H)$ is the Standard Model 
 operator as a function of the gauge invariant 
 bi-linear of the Higgs doublet. 
Once the Higgs doublet develops the VEV, 
 the tadpole term for the unparticle operator is induced,
\bea 
{\cal L}_{\slashed{\cal U}} =
 \Lambda_{\slashed{\cal U}}^{4-d_{{\cal U}}} {\cal U},  
\label{tadpole} 
\eea 
 and the conformal symmetry in the new physics sector is broken 
 \cite{U-Higgs}. 
Here, 
   $ \Lambda_{\slashed{\cal U}}^{4-d_{{\cal U}}}= 
   \langle {\cal O}_{\rm SM} \rangle/ \Lambda^{d_{\cal U}+n-4}$ 
 is the conformal symmetry breaking scale. 
At the same time, we have the interaction terms 
 between the unparticle and the physical Standard Model Higgs boson 
 ($h$) such as (up to ${\cal O}(1)$ coefficients) 
\bea 
  {\cal L}_{{\cal U}-{\rm Higgs}} 
  &=& \frac{\Lambda_{\slashed {\cal U}}^{4-d_{{\cal U}}}}{v} 
   {\cal U} h 
  + \frac{\Lambda_{\slashed {\cal U}}^{4-d_{{\cal U}}}}{v^2} 
    {\cal U} h^2
    \nonumber\\
  &+& \frac{\Lambda_{\slashed {\cal U}}^{4-2d_{{\cal U}}}}{v} 
   {\cal U}^2 h 
  + \frac{\Lambda_{\slashed {\cal U}}^{4-2 d_{{\cal U}}}}{v^2} 
    {\cal U}^2 h^2 + \cdots \;,   
 \label{mixing}
\eea 
where $v=246$ GeV is the Higgs VEV. 
In order not to cause a drastic change or instability 
 in the Higgs potential, 
 the scale of the conformal symmetry breaking  
 should naturally be smaller than the Higgs VEV, 
 $\Lambda_{\slashed {\cal U}} \lesssim v$. 
 When we define the `mass' of the unparticle as a coefficient of the second derivative
 of the Lagrangian with respect to the unparticle, ${\cal U}$,
then the mass of the unparticle can be obtained in the following form, 
$m_{\cal U}^{2-d_{{\cal U}}} = \Lambda_{\slashed {\cal U}}^{2-d_{{\cal U}}}$.

As operators between the unparticle and the Standard Model sector, 
 we consider 
\bea
{\cal L}_{\cal U} = 
 -\frac{\lambda_g}{4}  \frac{\cal U} {\Lambda^{d_{\cal U}}}
      G^A_{\mu \nu} G^{A \mu \nu} 
 -\frac{\lambda_\gamma}{4} \frac{\cal U}{\Lambda^{d_{\cal U}}} 
      F_{\mu \nu} F^{\mu \nu},  
\label{Unp-gauge} 
\eea
where we took into account of the two possible relative signs 
 of the coefficients, 
 $\lambda_g = \pm 1$ and $\lambda_\gamma= \pm 1$. 
We will see that these operators are the most important ones 
 relevant to the Higgs phenomenology.

As discussed before, once the Higgs doublet develops the VEV, 
 the conformal symmetry is broken in the new physics sector, 
 providing the tadpole term in Eq.~(\ref{tadpole}). 
Once such a tadpole term is induced, 
 the unparticle will subsequently develop the VEV
 \cite{U-Higgs, U-Higgs2} 
 whose order is naturally the same as the scale of 
 the conformal symmetry breaking, 
\bea
  \langle {\cal U} \rangle 
= \left( c \; \Lambda_{\slashed {\cal U}} \right)^{d_{{\cal U}}} . 
\eea 
Here we have introduced a numerical factor $c$, 
 which can be $c = {\cal O}(0.1) - {\cal O}(1)$, 
 depending on the naturalness criteria. 
Through this conformal symmetry breaking, 
 parameters in the model are severely constrained 
 by the current precision measurements. 
We follow the discussion in Ref.~\cite{U-Higgs}. 
 From Eq.~(\ref{Unp-gauge}), 
 the VEV of the unparticle leads to the modification of 
 the photon kinetic term, 
\bea
{\cal L} =
-\frac{1}{4}\left[ 
 1 \pm \frac{\langle {\cal U} \rangle}{\Lambda^{d_{{\cal U}}}}
\right] F_{\mu \nu} F^{\mu \nu} ,
\eea
 which can be interpreted as a threshold correction 
 in the gauge coupling evolution across the scale 
 $\langle {\cal U} \rangle^{1/d_{\cal U}}$. 
The evolution of the fine structure constant from 
 zero energy to the Z-pole is consistent 
 with the Standard Model prediction, 
 and the largest uncertainty arises 
 from the fine structure constant 
 measured at the Z-pole \cite{Yao:2006px}, 
\bea
 \widehat{\alpha}^{-1}(M_Z) &=&  127.918 \pm 0.019 . 
\nonumber 
\eea
This uncertainty (in the $\overline{\rm MS}$ scheme) 
 can be converted to the constraint,  
\bea
\epsilon=\frac{
\left<{\cal U} \right>}{\Lambda^{d_{{\cal U}}}}
 \lesssim 1.4 \times 10^{-4}. 
\label{gauge}
\eea
This provides a lower bound on the effective cutoff scale. 
For $d_{\cal U} \simeq 1$ and $\Lambda_{\slashed{\cal U}} \simeq v$ 
 we find 
\bea
\Lambda \gtrsim c \times 1000~{\rm TeV} ,  
 \nonumber 
\eea
This is a very severe constraint on the scale of new physics, 
 for example, $\Lambda \gtrsim 100$ TeV for $c \gtrsim 0.1$.

\section{Supersymmetic unparticle}
A supersymmetic extension of the original unparticle physics has been
proposed by \cite{U-SUSY}. We begin by reviewing the scenario of the supersymmetic unparticle
and give something more details.

The unparticle which has originally been introduced as a sort of scalar operator is now extended to be a chiral multiplet
in order to fit with the supersymmetic theory. Explicitly speaking, it is written as
\be
{\cal U} = {\cal U} + \sqrt{2} \theta_\alpha {\cal U}^\alpha + \theta^2 F_{\cal U} \;.
\ee
Here, $\alpha$ is a spinor index, and we write the same notation for the scalar component of the unparticle 
chiral multiplet as the chiral multiplet itself.
Then the interaction or the superpotential between the unparticle and the MSSM sector is, in general, given in the same way
as the non-supersymmetic unparticle:
\be
{\cal L} =  \int d^2 \theta  \frac{1}{\Lambda^{d_{\cal U}+ n -3}} \, {\cal U} {\cal O}_{\rm MSSM} + h.c. 
\ee

 \subsection{SUSY QCD as a natural candidate of the unparticle}
The most promising example of the SUSY unparticle is given by the SUSY QCD based on $SU(N_c)$ gauge symmetry
with $N_f$ flavors \cite{Seiberg:1994bz}, which is a natural SUSY extension of the Banks-Zaks model \cite{Banks:1981nn}. 
This correspondence has already been noted in the literature \cite{U-Higgs}.
We take $\frac{3}{2}N_c \le N_f \le 3N_c$ so that the unparticle SUSY QCD is in the conformal window. 
We denote the chiral superfields for the $N_f$ flavors by $Q_i$ and $\bar{Q}_{j}$ ($i,j = 1\dots N_f$). 
$Q_i$ transforms as a fundamental representation of $SU(N_c)$ and $\bar{Q}_j$ transforms as an anti-fundamental representation.
In general, SUSY QCD in the conformal window ($\frac{3}{2}N_c \le N_f \le 3N_c$)
flows to a strongly coupled conformal fixed point in the infrared (IR).
At the fixed point the theory has a dual description (Seiberg dual) with a gauge group $SU(N_f-N_C)$,
$N_f$ dual-quark superfields ($Q$, $\bar{Q}$), a gauge singlet meson superfield $M_{ij}$ 
(transforming in the bifundamental representation of the $SU(N_f) \times SU(N_f)$ flavor symmetry, 
and the superpotential is given by
\be
W=\bar{Q}^i M_{ij} Q^j \,.
\ee
The meson superfield $M_{ij}$ in the dual description corresponds to the gauge invariant composite $\bar{Q} Q$ of the original theory.
In regards to the unparticle physics, SUSY QCD allows us to determine some parameters in an explicit way. 
The conformal dimension of the meson superfield $M_{ij}$ is fixed by the $R-$symmetry:
\be
d_{\rm UV} = d_M = 3 \frac{N_c - N_f}{N_f}  \;.
\ee
It has to be noted that $1 \le  d_{\rm UV} \le 2$  in the conformal window ($\frac{3}{2}N_c \le N_f \le 3N_c$).

\subsection{A solution to the $\mu$ problem}
Now, we can go to the discussion of the unparticle physics in the Higgs sector. 
Providing a coupling with Higgs sector for the unparticle with non-vanishing VEV 
can provide a natural solution to the $\mu$ problem in the MSSM 
exactly the same way as in the Next to minimal supersymmetric Standard Model (NMSSM).
Given the following superpotential for the unparticle in connection with the Higgs sector:
\bea
{\cal L} &=& k_\mu \int d^4 \theta  \, \frac{1}{\Lambda^{d_{\cal U} }} {\cal U}^\dag  H_u H_d +h.c.
\nonumber\\
& \to &  \int d^2 \theta \, \mu  H_u H_d \;.
\eea
Here, the $\mu$ term is generated when the unparticle develops the VEV,
\be
\mu = k_\mu \frac{F_{\cal U}}{\Lambda^{d_{\cal U}}} \;. 
\ee
This provide a viable and interesting solution to the $\mu$ problem in the MSSM
since the origin of the VEV of unparticle is related to the conformal symmetry breaking
and the scale of it could be as much as higher than the weak scale in contrast to
the NMSSM cases.

Interesting point in our scenario of unparticle physics as a source of  $\mu$ term
is that the overall scale of a supersymmetric mass parameter $\mu$ is determined
by the scale of the conformal symmetry breaking and the dimension of the unparticle operator, $d_{\cal U}$.
For instance, if we take $k_\mu \sim {\cal O}(1)$ and $\Lambda = 10$ TeV in order for obtaining $\mu \sim 100$ GeV,
the required value of the unparticle $F$-term is of order, $\sqrt{F_{\cal U}} \sim 1$ TeV for $d_{\cal U} =1$
and $F_{\cal U}^{1/3} \sim 10$ TeV for $d_{\cal U} \sim 2$,
which are relatively low compared to the case of usual gravity mediation since the cutoff scale in this case
becomes Planck scale.

On the other hand, generating a sizable $B_\mu$ term is a on the same footing problem
as the $\mu$ problem in the MSSM. In our scenario, in which the $\mu$ term is generated
via the conformal symmetry breaking VEV of the unparticle, the $B_\mu$ term can be written 
in the following manner:
\bea
{\cal L} &=&  k_B \int d^4 \theta \, \frac{1}{\Lambda^{2 d_{\cal U} }} {\cal U}^\dag {\cal U} H_u H_d +h.c.
\nonumber\\
&\to &
B_\mu = k_B \frac{\left| F_{\cal U} \right|^2}{\Lambda^{2 d_{\cal U} }} \;. 
\eea
Natural electroweak symmetry breaking requires the scale of $B_\mu$ term of order, 
$B_\mu \sim \mu^2$. This is indeed true in our scenario of unparticle physics 
as a source of both $\mu$ and $B_\mu$ terms if $k_\mu \sim k_B$. 
It is always the case in minimal supergravity with additional singlet, 
which is used to generate $\mu$ term, as in the Giudice-Masiero mechanism.

\section{Summary}
In conclusion, we have considered the unparticle physics 
 focusing on the Higgs phenomenology. 
Considering the interactions between the unparticle and Higgs boson, unparticle can acquires
the VEV, whose mass scale is determined as a consequence of the conformal symmetry breaking.
The result shows that there is a possibility for the unparticle as a hidden sector in
SUSY breaking sector, and can provide a solution to the $\mu$ problem
in SUSY models.

\begin{center}
{\bf Acknowledgments}
\end{center}
We would like to thank N. Okada for his stimulating discussions.
The work of T.K. was supported by the Research
Fellowship of the Japan Society for the Promotion of Science (\#1911329).


\end{document}